\documentclass[1p,11pt,sort&compress]{elsarticle}
\usepackage{etoolbox}
\makeatletter
\def\ps@pprintTitle{%
 \let\@oddhead\@empty
 \let\@evenhead\@empty
 \def\@oddfoot{\rightline{\it{\today}}}%
 \let\@evenfoot\@oddfoot}
\makeatother
\usepackage{lscape}

\usepackage{epsfig}
\usepackage{url}

\usepackage{amssymb}
\usepackage{geometry}

\usepackage{amstext}    % defines the \text command, needed here
\usepackage{array}      % for advanced column specification: use >{\command} for
                        \usepackage{amsmath,tabu}

\usepackage{graphicx}

\usepackage{epstopdf}

\usepackage{amsmath}

\usepackage[makeroom]{cancel}

\usepackage{mathpazo}

\usepackage{slashed}

\usepackage{mathtools}
\usepackage{ulem}

\newcommand{\ignore}[1]{}
\newcommand{\be}{\begin{equation}} \newcommand{\ee}{\end{equation}}

\newcommand{\ab}{\allowbreak}

\def\ba#1\ea{\begin{align}#1\end{align}}
\newcommand{\bit}{\begin{itemize}}
\newcommand{\eit}{\end{itemize}}

\newcommand{\Rdot}{R_{\infty \bullet}}

\newcommand{\nn}{\nonumber} \renewcommand{\bf}{\textbf}
\newcommand{\ra}{\rightarrow}

\newcommand{\R}{R_{\infty}}

\renewcommand{\a}{\alpha}

\newcommand{\p}{\partial}

\def\slashb#1{\setbox0=\hbox{$#1$}#1\hskip-\wd0\dimen0=5pt\advance
        \dimen0 by-\ht0\advance\dimen0 by\dp0\lower0.5\dimen0\hbox
          to\wd0{\hss\sl/\/\hss}}
\input{epsf}
\usepackage[modulo]{lineno}
\journal{Physics Letters B}
\makeatletter
\renewcommand{\@makefntext}[1]
{\noindent\makebox[1.5em][r]{\@makefnmark}\hspace*{.3em}%
\addtolength{\linewidth}{-1.8em}%% (this is local)
\parbox[t]{\linewidth}{#1}}
\makeatother

\begin{document}

\begin{frontmatter}



%\newcounter{tempfootnote}
%\setcounter{tempfootnote}{\value{footnote}}
%\setcounter{footnote}{1}
%\renewcommand{\thefootnote}{\fnsymbol{footnote}}
%\author{John Craig Martens\corref{cor1}\fnref{label2}\footnote{Corresponding author. {\\\it Email address: j671m042@ku.edu} %(J.C. Martens)}}
%\author{John P. Ralston\corref{cor1}\fnref{label2}}

%\newcounter{tempfootnote}
%\setcounter{tempfootnote}{\value{footnote}}
%\setcounter{footnote}{1}
%\renewcommand{\thefootnote}{\fnsymbol{footnote}}
\author{John Craig Martens\corref{cor1} and John P. Ralston\corref{}}
\ead{martens@ku.edu}
\cortext[cor1]{Corresponding author}

 %%\ead{email address}
%\fntext[label4]{Corresponding author. {\it Email address: j671m042@ku.edu} (J.C. Martens)}
%%\cortext[cor1]{}

\title{The Muon Experimental Anomalies  \\ Are Explained by a New Interaction \\ Proportional to Charge\vspace{-1ex}}

%%\author[label1,label2]{}
%%\address[label1]{}
\address{Department of Physics and Astronomy, The University of Kansas, Lawrence, KS 66045, USA\vspace{-5ex}}
%\author{}
\begin{abstract} The ``proton size puzzle'' and the ``muon anomalous moment problem'' are incomplete descriptions of significant discrepancies of Standard Model calculations with experiments. What is particularly new is that the experiments and theory confront a new regime of ultra-precise physics where traditional piece-meal analysis methods fail to be self-consistent. At current levels of precision the proton size $r_{p}$, the Rydberg constant $\R$, the fine structure constant $\a$ and the electron mass (Compton wavelength $\lambda_{c}$) are inextricably coupled, so that the actual discrepancies might be almost anywhere, while merely {\it appearing} to be muon-derived through a historical order of assumptions. We have conducted a new global fit to all of the relevant data using the entire body of Standard Model theory. A conventional $\chi^{2}$ statistic is used to fit all relevant fundamental constants with and without a generic ``no-name'' boson of undetermined spin that interacts universally with leptons and hadrons proportional to electric charge. The analysis discovers a new local minimum region of $\chi^{2}$ where all of $r_{p}, \, \R, \, \a, \, \lambda_{e}$ have new values compared to previous work, while accommodating all of the data, unlike previous determinations. A new particle $X$, possibly related to the ``dark photon'' but more generally defined, is predicted to be observed in electron- and muon-based experiments. 

\end{abstract}

\end{frontmatter}

\section{More Than One Interconnected Experimental Anomaly}

High precision experiments on muons disagree with Standard Model predictions. The muon magnetic moment parameter $a_{\mu}$ measured at Brookhaven National Lab\cite{Bennett:2004pv} is larger than calculations by $2.9 \times 10^{-9}$, a discrepancy reported variously in the literature as $4.6 \sigma$ or larger. The $2S_{1/2}2P_{1/2}$ Lamb shift measured in muonic hydrogen by the CREMA collaboration at PSI differs by 75 GHz from atomic QED calculations\cite{Pohl:2013yb} a discrepancy reported as exceeding $7\sigma$, based on comparison of the charge radius parameter from published tables. Fermilab will re-measure $a_{\mu}$ in the near future. The CREMA collaboration should soon release new measurements, including the muonic deuterium Lamb shift. A new interdisciplinary community combining atomic, nuclear, and particle physics expertise is uniting to confront the discrepancies.\cite{Pohl:2013yb} The new community is optimistic that new ultra-precise muon-proton(\cite{Gilman:2013eiv, 2013AIPC.1563..167G}, and electron-proton scattering experiments can be conducted and add new information at the cutting-edge of technological feasibility. New muon-specific interactions\cite{Brax:2010gp, Barger:2010aj, TuckerSmith:2010ra, Batell:2011qq, Barger:2011mt, Karshenboim:2014tka, Wang:2013fma, Li:2013dwa, Brax:2014zba, Carlson:2015jba} have been proposed to explain the experimental discrepancies, while giving up lepton universality has a high cost.

In contrast, low energy electron observables agree exceedingly well with electroweak theory. Atomic QED theory has made the Rydberg constant of electronic hydrogen ``the most precisely determined physical quantity''. That statement, however, assumes QED and weak theory are correct, which the muonic data contradict. The electron magnetic moment calculated in electroweak theory so precisely agrees with experiment it has become the defining standard of the fine structure constant. This turns out to be an issue. The dominant conclusion has been that any new universal interaction, of sufficient size to explain the muon data, would produce much more visible effects with electrons, causing discrepancies not observed. Yet giving up interactions with electrons terminates a wide spectrum of new observables that might resolve the discrepancies.

Actually what is computed for electrons depends on fundamental constants. When it comes to electrons the constants have an unrecognized danger of circularly confirming what is measured. The superb agreement of theory and experiment for the electron's anomalous moment parameter $a_{e}$ does not itself test anything. That is because these quantities became {\it de facto} definitions of the fine structure constant $\a$ once the experimental and theoretical uncertainties became much smaller than all other measures. {\it IF} QED and weak theory are exact, the highest precision data and theory produce the highest precision constants. Yet an unknown interaction might contribute to $a_{e}$ and shift {\it the value ascribed to} $\a$ in an utterly undetectable way. The actual tests come from comparing independent observations that are not circular. Tests involving $\R$ actually depend on $\a$, the proton charge radius, and so on: There are no fundamental physical constants that are {\it not} coupled to other fundamental constants.

We come quickly to a new space where the value and uncertainty of a fundamental constant cannot reliably be found in government-approved tables. That is one reason for reading abundant warnings found with the tables. {\it IF} there is other physics at work, the nominal precision of a constant fit to an incomplete theory can be pure illusion. The main reason for physicists to care about high-precision constants is to spot discrepancies and find new physics. But upon making the hypothesis that new physics is relevant, the constants and their uncertainties from the previous hypothesis cease to be reliable guideposts.  

The meaning and the uncertainty of {\it all} constants depends on the hypothesis. The proton charge radius has an unchanging theoretical definition $r_{p}^{2} =-(1/6)\p G_{E}/\p q^{2}$ evaluated at momentum transfer-squared $q^{2}=0$, where $G_{E}$ is the Sachs electromagnetic form factor.  Dozens of measurements of $q^{2}$ dependence have reported estimates of $r_{p}^{2}$ found from extrapolation to $q^{2}=0$. Making an extrapolation is self-consistent {\it under the hypothesis} the form factor is dominated by known hadronic singularities in the complex $q^{2}$ plane. Yet recently the uncertainties of extrapolation have gotten more attention and become controversial for the proton size puzzle\cite{Bernauer:2013tpr, Lee:2015jqa, Higinbotham:2015rja}. It has not been noticed that extrapolations become unreliable under a different hypothesis that a new, sufficiently low-mass weak interaction might exist. In that case the complex plane singularities of a new scattering amplitude could be so close to $q^{2} \sim 0$ they might be unobservable, besides lying outside the conceptual universe of hadronic fits. Thus the {\it ``experimentally-derived''} meaning of ``proton charge radius'' and its error bars actually depend critically on the theory used to interpret data. The situation with electron and muon scattering, then, is even less settled than perceived, while still demanding more experimental study in any scenario. 

The perception that the proton size has been precisely and unconditionally determined in electronic hydrogen ($eH$) spectroscopy is also flawed. The spectroscopic data actually determines a {\it correlation} between two free parameters, which are $r_{p-eH}$ and the Rydberg constant $R_{\infty}$. The correlation coefficient is 0.99, meaning that $r_{p-eH}$ and $R_{\infty}$ can be varied quite a bit along a straight line while giving a good fit. It is a basic concept error to use error bars without attending to the correlations. Thus the notion that $R_{\infty}$ and its cited uncertainty could be used in isolation to constrain new physics effects lacks a self-consistent foundation. But there is more. The $eH$ spectroscopic fits are done with the values of $\a$ and the electron-proton mass ratio $m_{e}/m_{p}$ {\it fixed} by other experiments. If $\a$ or $m_{e}/m_{p}$ are varied, the values of $r_{p-eH}$ and $\R$ can easily vary well outside their nominal uncertainties.  

Miller {\it et al}\cite{Miller:2011yw} forcefully emphasized that ``the proton size puzzle'' signals something deeply wrong with current physics, and something not to be shoved aside as an unimportant parameter detail. We agree, and enlarge the scope to discover a global question that cannot be resolved by piecemeal methods. The question itself is challenging: How could anyone think they understand how
the proton size, the Rydberg, the fine structure constant (and then) the electron mass are so inextricably coupled, to know for sure the actual discrepancies is in muons? The discrepancies are so small and so subtle they might {\it appear} to be muon-based, simply due to a historical order of analysis and circular assumptions. 

To proceed we have conducted a global fit to all of the relevant data using the entire body of Standard Model theory. We compare fits to the data with and without a generic ``no-name'' boson of undetermined spin that interacts proportional to electric charge. To the extent they might apply, we review exclusion limits developed for ``dark photons,'' a highly specific model of great current interest\cite{Fayet:2006sp, Pospelov:2008zw, Essig:2013lka}. Our model depends on two parameters $\a_{X}$, which is a dimensionless coupling analogous to $\a$, and the boson mass $m_{X}$. In the region of $m_{X} \gtrsim 50$ Mev the analysis depends only on the parameter combination $\a_{X}/m_{X}^{2}$. The null model is $\a_{X} \ra 0$ with $\a, \, R_{\infty}, \, r_{p}, \, \lambda_{c}$ as fitting parameters. We conduct a simple $\chi^{2}$ hypothesis test which compares the null (Standard) model with the model fitting $\a_{X}$. The null model is ruled out by more than 15 units
of $\chi^{2}$ of $m_{X}>50$ MeV. The improvement in fit is more than possible fitting $a_{\mu}$ and $r_{p-\mu}$ alone with a muon-specific interaction, while also using fewer parameters. The model coupling proportionately to electric charge is not strictly required, but a very small coupling to neutrons is certainly needed. 

Put differently, the fits to electronic and muonic hydrogen are in principle capable of predicting electronic and muonic deuterium\cite{Krauth:2015nja} with no free parameters. The model passes the test with electronic deuterium data that exists, which can be tested when muonic deuterium becomes available.

The statistics are robust and unchanged by deleting different types of data. No particular subset of data dominates, meaning that $\chi^{2}$ is acceptable for each type of electron or muon magnetic moment, hydrogen, or deuterium data.  Fits consistently find the actual proton charge radius $r_{p} \sim 0.84$ fm, which is close to the one found by the muonic Lamb shift. We call this the {\it ``minimal-universal  solution,''} which is completely unexpected. The only cases indicating the muonic charge radius differs significantly from the true one are those excluding the muonic Lamb shift data entirely. 

Electronic deuterium ($eD$) spectroscopy provides a highly non-trivial test. The minimal-universal  solution plus nuclear theory predicts the deuteron charge radius with no free parameters. The model predicts the charge radius of the muonic deuterium Lamb shift eagerly awaited from the CREMA collaboration measurements.  The most surprising aspect of the small proton solution concerns electronic hydrogen spectroscopy. The combination of our value of $\a$, the correlation of $\R$ and $r_{p-eH}$, and the value of $\a_{X}$ produces a substantial revision of the Rydberg constant, while greatly improving the global fit compared to the Standard Model. The best fit parameters are shown in Table 1 in the next section.

\section{Data and Fitting Procedure}

\subsection{Observables}

The experimental observables of the analysis are: \ba & a_{e} =0.00115965218073  \pm 2.8 \times 10^{-13 }  \nn \\   
& a_{\mu} = 0.00116592091 \pm 6.3 \times 10^{-10} \nn \\ 
& \mu H: \quad \Delta E_{2S-2P} =  202.3706 \pm 0.0026 \, \text{meV}   \nn \\ 
& m_{e}/h = 0.7634407125716617609 \times 10^{20}  \, \text{MeV}\nn \\ 
&  e H: \text{ 7 transitions listed in Table \ref{tab:transitions}} \nn \\
&  e D: \text{ 7 transitions listed in Table \ref{tab:transitions}} \nn\ea  Here $\mu H$ stands for muonic hydrogen, while (to repeat) $eH$ and $eD$ stand for electronic hydrogen and deuterium. We accept $ {m_{e} /m_{p}}=1836.152672444 $, $m_{D}/m_{e}=3670.48296513$, $m_{\mu} =106.7$ MeV as given values. We express $\a$ and $\R$ in units of reference values $ \a_{{\bullet}} =0.0072973525664, \, \Rdot =10973731.5685080 \,\text{m}^{-1}, \, \lambda_{C \bullet} = 2.4263102367(11) \times 10^{-12} \, \text{m}$. Except for definitions, nothing in our analysis depends on these numbers.

\subsection{Procedure}

Our analysis fits a conventional $\chi^{2}$ statistic \ba  &  \chi^{2}= \sum_{j} \,  {(d_{j}-t_{j}(\theta_{\ell}) )^{2} \over \sigma_{j}^{2}}. \label{chidef} \ea  Here $d_{j}$ and $t_{j}$ stand for the $j$th instances of data, and theory respectively, with experimental uncertainty $\sigma_{j}$. Fitted parameters are $\theta_{j} =(\a, \, \R , \,r_{p},  \,\theta_{X})$, where new physics parameters are $\theta_{X} =(\a_{X}, \, m_{X})$ (low mass) or the combination $\xi=\a_{X}/m_{X}^{2}$ (high mass). Fits respect the defining relation $\R= \a^{2} /2\lambda_{c}$, where $\lambda_{c}=h/m_{e}=0.00072738950972 (34) m^2/s$ is the electron Compton radius measured experimentally\footnote{Neither $m_{e}$ nor $h$ has been determined with the precision needed in the study, and neither value appears anywhere in the analysis. Their uncertainties are 100\% correlated, and cancel in $m_{e}/h$, sometimes called the quantum of circulation, which we express with $\lambda_{c}$.} as $m_{e}/h$. In full detail \ba \chi^2= &{(a_e^{exp}-a_e^{theory}(\a, \, \theta_{X}))^2 \over \sigma^2(a_e)} \nn \\
& +{(a_{\mu}^{exp}-a_{\mu}^{theory}(\a, \, \theta_{X}))^2 \over \sigma^2(a_{\mu})} \nn \\ 
& + \sum_j^8 \, {(\Delta f_{eH,j}^{exp}-\Delta f_{eH,j}^{theory}(\a,\, \R, \, r_{p}, \, \theta_{X}))^2 \over \sigma^2(\Delta f_{eH}) }\nn \\ 
& + \sum_j^8 \, {(\Delta f_{eD,j}^{exp}-\Delta f_{eD,j}^{theory}(\a,\, \R, \, r_{p}, \, \theta_{X}))^2 \over \sigma^2(\Delta f_{eD})} \nn \\
&  + {(\Delta f_{\mu H}^{exp}-\Delta f_{\mu H}^{theory}(r_{p}, \,  \theta_{X}))^2 \over \sigma^2(\Delta f_{\mu H})} \nn \\
& + { (4  \pi c R_{\infty}/\alpha^2 -   (m_e/h)^{exp}) \over \sigma^2(m_e/h) } \nn \\
\ea The terms in the order shown will be called $\chi^{2}(a_{e}), \ab \, \chi^{2}(a_{\mu}), \, \ab \chi^{2}(eH),\ab  \, ~\chi^{2}(eD), \ab \, \chi^{2}(\mu H), \, \chi^{2}(\lambda_{c})$ when discussed separately. The parameters we vary are displayed explicitly in the expression above, while others whose variation is safely suppressed are set to the reference values. For example $a_{e}$ is exquisitely sensitive to $\a$ and $\a_{X}$. In QED-electroweak (QED-EW) theory $eH$ and $eD$ have long been fit with two parameters $\R$, $r_{p}$, setting $\a=\a_{\bullet}$ obtained from $a_{e}$. When $\a_{X}$ is included $\R$ is highly sensitive, and $\a$ must be included in the $eH$, $eD$ fits for self-consistency of the definition of $\R$. In comparison the $\a$ dependence of $f_{\mu H}^{theory}(r_{p}, \,  \theta_{X}))$ over the range of interest is too weak to matter. These facts were determined before the analysis using estimated parameter uncertainties and checked after the analysis. Discussion of a procedure including additive parameters for systematic theory uncertainty is given in the Appendix.

The bound state lepton-proton effective potential $V_{X}$ is \ba V(x) = \a_{X}{e^{-m_{X}r} \over 4 \pi r}. \nn \ea The anomalous moment calculations are done with the relativistic Lagrangian density of fermions minimally coupled to a massive vector field with the same $\a_{X}$ coupling at one-loop order\cite{Fayet:2007ua}. Since no other details about the theory are needed or can be observed in our analysis, the no-name boson model has not been restricted to a particular Lagrangian.

There are two important facts about the couplings. Finding the new $\chi^{2}$ minimum was complicated by the
fact the minimum $\chi^{2}$ region is tube-shaped in a five-dimensional parameter space. For $m_{X}$ larger than the muon mass a degenerate dependence on the combination $ \xi = \a_{X}/m_{X}^{2}$ is expected. At small mass the nature of the best-fit region is new and comes from the interplay of all the coupling constants. The minimum value of $\chi^{2}$ falls rapidly by more than 10 units between $m_{X}=10$ MeV and $m_{X} =50$ MeV, where the difference $\Delta \chi^{2}=12$. Arbitrarily choosing $\Delta \chi^{2}=4$ to define significance, the entire region of $m_{X} \gtrsim 20$ MeV is favored. Once $m_{X}$ is large enough for significance, judgment with other information is needed to decide a preferred value of $m_{X}$, which we leave undetermined. The minimum value of $\chi^{2}$ continues to fall monotonically as $m_{X}$ increases, but at a decreasing rate, reaching $\Delta \chi^{2}>15$ for $m_{X}\gtrsim$ GeV and above. Since the best fit parameters are found along a curve in the $(\a_{X}, \, m_{X})$ plane, there is only one controlling variable in the favored region. Expressing fits in terms of the controlling variable combination would explicitly remove one new parameter from the analysis. For simplicity we report results in terms of $\a_{X}$ while stepping through values of $m_{X}$, which superficially appear to be completely independent parameters, although they are not.

Continuing, the sign of lepton couplings is unobservable in the anomalous moments, and in many other observables. The bound state $eH$, $eD$ and $\mu H$ data depend critically on the sign of the coupling. The smaller value of $r_{p}$ extracted from the muonic Lamb shift, compared to QED-EW fits of $eH$, had previously led to a widespread conclusion that any new lepton-proton interaction must be attractive. We left the sign undetermined, and discovered the new local minimum region with a {\it repulsive} interaction ($\a_{X}>0$). The sign of $\a_{X}$ is what makes the apparent (QED-only) fitted value of $r_{p-eH}$ larger than the size found in $\mu-H$, exactly as the data goes. 

In addition, the relatively larger effects in $eH$, generally considered a barrier to a universal lepton interaction, turned out to be crucial in the final fit. An electron interaction moving $r_{p-eH}$ from a true value near 0.84 (henceforth in fm units) to a QED-fit value of 0.88 in $eH$ is much smaller than a muon-specific coupling adjusting the muonic Lamb shift the other way. The new interaction is so small, and the momentum transfer so small, that the muonic $r_{p -\mu H}$ is very close to the true one. 

Then a smaller value of $\a_{X}$ diminishes the potential tension of fitting $a_{e}$ while simultaneously fitting $a_{\mu}$. The fitted value of $a_{e}$ is also partly compensated by a small change in $\a$ that would be unobservable in $a_{e}$ alone. The final critical element for determining $\a$ is the electron mass ratio $m_{e}/h$, expressed in our fits with the Compton wavelength $\lambda_{c}=h/m_{e}c$. The experimental value was obtained by combining the Rubidium to electron mass ratio $m_{Rb}/m_{e}$ with Rubidium recoil measurements\cite{2011PhRvL.106h0801B,  2013AnP...525..484B} of $m_{Rb}/h$. This last fact -- that a single Rubidium experiment dominates the actual tests of the electron anomalous moment theory -- is the same as in QED-EW theory, and has been noted before\cite{Aoyama:2014sxa}.

For convenience of readers, Table 1 shows the changes of our best-fit parameters $\delta \theta_{i}$ relative to the QED-EW based CODATA2014 values\cite{Mohr:2015ccw}. It is easy to check that $\delta \a/\a_{\bullet}$ happens to be well within the uncertainty permitted by the QED-EW fit. That is both fortuitous and logically unnecessary. The Rydberg happens to be revised with $\delta \R/\Rdot$ about 2-3$\sigma$ different from recent compendia. This is irrelevant, because the previous cross-check on $\R$ and $\a$ are the same uncertainty of the electron mass ratio $m_{e}/h$ that we fit as well or better. Our overall fit is actually much better than previous ones in two ways. First, a significantly lower value of $\chi^{2}$ is obtained, even accounting for more parameters. (Wilks' theorem predicts that if the null fits the data, then a model adding one extra parameter and smoothly connected with the null will have $\Delta \chi^{2}$  distributed by $\chi_{1}^{2}$. More concretely, whenever $\Delta \chi^{2} \geq few$ when adding one parameter the null is in danger of being ruled out.) Second, our fit is the first high-precision fit to fundamental constants using $a_{e}$, $eH$ and $eD$ data that does not throw out either $a_{\mu}$, $\mu H$, or both.

\subsection{Results}

\begin{figure}[ht]
\begin{center}
\includegraphics[width=5in]{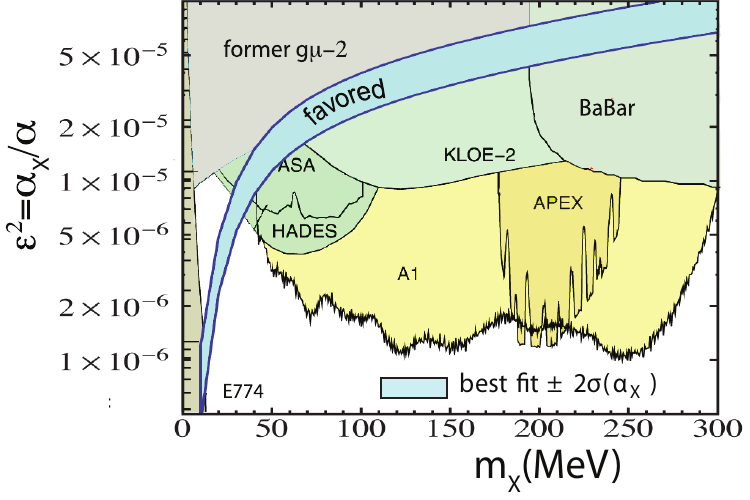}
\caption{A partial view of the region of the $(\a_{X}, \, m_{X})$ plane favored by the analysis. The statistic $\Delta \chi^{2}>4$ for $m_{X}>20$ MeV drops rapidly to $\Delta \chi^{2}>13$ for $m_{X}>50 $ MeV, and then decreases monotonically at a much slower rate for larger $m_{X}$. No upper limit on $m_{X}$ can be resolved, with $\Delta \chi^{2}$ reaching 15 in the GeV region and above. Favored region is found from the minimum $\chi^{2}$ curve varied with $\pm 2$ units of the standard uncertainty of $\a_{X}$. For informational purposes, regions where dark photons are weakly excluded by previous experiments are also shown. Extra assumptions and conditions for those limits and symbol $\epsilon^{2} \ra \a_{X}$ to apply to the analysis are reviewed in Section \ref{sec:limits}.}

\label{fig:limitplot}
\end{center}
\end{figure}

A concise summary of our results is shown in Figure \ref{fig:limitplot}. Over the region labeled ``favored'', we fit all the data with a high statistical significance. The difference of best fit values ranges from $\Delta \chi^{2}>4$ (left edge, $m_{X}>20$) falling rapidly to $\Delta \chi^{2}>13$ at $m_{X}=50$ MeV.  Then $\Delta \chi^{2}$ slowly decreases as $m_{X}$ increases The relation between $\epsilon^{2}$ and $\a_{X}/\a$ symbols (which do not generally have the same meaning) is explained in Section \ref{sec:limits}.

Table \ref{tab:parameterTable} shows the results of several fits adjusting $m_{X}$. The standard uncertainties found from the  inverse parameter covariance matrix are given by the values in parentheses. The dependence on $m_{X}$ is smooth and values at intermediate points can be inferred from the points shown. For $m_{X} \gtrsim many$ GeV the analysis remains consistent, but we expect exclusion limits to become severe.

\begin{figure}[ht]
\begin{center}
\includegraphics[width=3.8in]{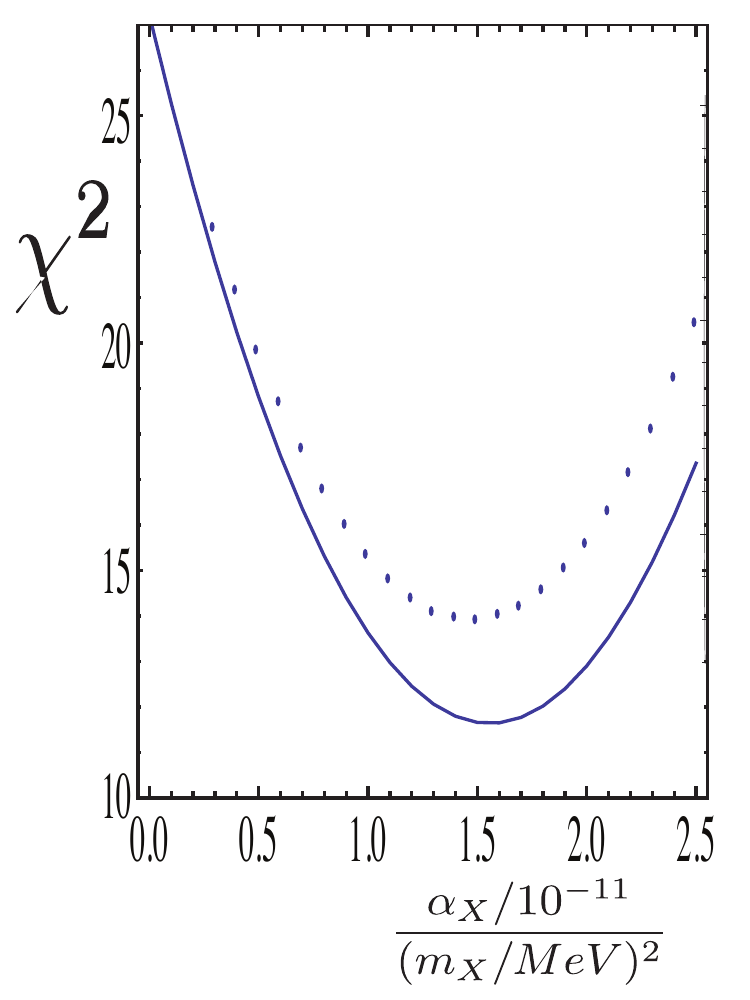}
\caption{The value of $\chi^{2}$ of the full data set as a function of $\a_{X}$ evaluated at the arbitrary value $m_{X}=50 $MeV. Parameters not shown are evaluated point by point as $\a_{X}$ is varied. The top (bottom) curves use $\chi^{2}$ as defined in the text, and with experimental uncertainties doubled, respectively. }
\label{fig:Chi2Comp}
\end{center}
\end{figure}

\begin{table}[ht]
	\centering
\resizebox{.83\textwidth}{!}{\begin{minipage}{\textwidth}
\begin{math}
\begin{array}{ccccccc}
 \hline
 \text{Omit} & \chi^2_{tot} & \Delta \chi^2 &
    (\delta  \R/\R^{\bullet})/10^{-12} & (\delta  \alpha/\alpha^{\bullet})/10^{-10} & r_p \, [fm] &
   \xi /10^{-11} \\
   \hline
   \hline
 \text{none} & 14.3 & 13.5 & 610(430) & -3.1(2.1) & 0.84113(27) & 1.40(38) \\
 \text{$\lambda_c$} & 11.0 & 16.1 & 1290(910) & -6.5(4.4) & 0.84117(27) & 1.60(43) \\
 \text{$\mu$H} & 10.1 & 13.0 & 620(410) & -3.1(2.1) & 0.88143(27) & 1.39(38) \\
 \text{$a_{e}$} & 11.0 & 16.1 & -17(12) & 0.014(10) & 0.84117(27) & 1.60(43) \\
 \text{$a_{\mu}$} & 11.7 & 0.3 & 60(42) & -0.38(26) & 0.84074(27) & -0.81(22) \\
 \text{$a_{e}$,$a_{\mu}$} & 6.9 & 4.6 & -8.3(5.9) & 0.058(40) & 0.84650(27) & 31.5(8.6) \\
 \text{$\mu$H, $a_{\mu}$} & 6.9 & 0.6 & -8.3(5.9) & 0.058(40) & 0.88453(27) & -1.14(30) \\
 \text{eH} & 7.4 & 13.1 & 610(430) & -3.1(2.1) & 0.84112(27) & 1.39(38) \\
 \text{eD} & 10.0 & 13.4 & 610(430) & -3.1(2.1) & 0.84113(27) & 1.40(38) \\
 \text{eH, eD} & 0.0 & 15.7 & -1310(920) &
   -6.5(4.4) & 0.84116(27) & 1.57(43) \\
   \hline
   \hline
\end{array}
\end{math}

\end{minipage}}
\caption{The parameters for the best fit to all the data and for fits where observables are removed. Parentheses list the standard uncertainties. $\Delta \chi^{2}$ is the difference of $\chi^{2}$ of the null model ($\a_{X}=0$) with the best fit.  Table is made with an arbitrary value of $m_{X}=50$ MeV.}

\label{tab:parameterTable}

\end{table}

Table \ref{tab:DeutoCompo} compares the full fit to those removing deuterium spectroscopic data. As a rule the deuterium data changes parameters by very little. This is significant because the deuterium charge radius is predicted from $r_{p}$, which by nearly coinciding with the $\mu H$ determination, changes $r_{d}$ very significantly compared to QED-EW fits.

\section{Theory and Code Validation} 

We validated our implementation of QED-EW theory extensively by comparing it to previous work. While space limitations preclude listing all formulas, a summary of how our work was undertaken will be given. 

\subsection{Analysis Overview} 

\subsubsection{Moments}

For example, in the region of $m_{X} >> m_{e}$, the Standard Model\cite{Aoyama:2014sxa} plus one-loop contribution of new physics to the electron anomalous moment is summarized by \ba  & a_{e}^{theory} = 1.7147 \times 10^{-12} +0.159155 \alpha
   -0.0332818 \alpha ^2  +0.0380966
   \alpha ^3  \nn \\ &
 -0.0196046 \alpha ^4 + 0.0299202  \alpha ^5
+0.027706 \, \xi m_{X}^{2}f(m_{X}/m_{\ell}),\nn \\ &  \qquad \text{where} \quad \xi= { \alpha_{ X} \over m_{X}^2 }. \label{aeform} \ea Here $m_{\ell}$ is the lepton mass and $f(m_{X}/m_{\ell})$ is an integral expression from the one-loop calculation found in the literature\cite{Leveille:1977rc}. In the limit $m_{X}/m_{\ell}>>1$ then $f \ra 1$. So long as the experimental and theoretical uncertainties of $a_{e}$ are sufficiently small compared to other observables, this formula becomes a {\it de facto} definition of $\a$. It also exposes the degeneracy of determining $\a$ in conjunction with parameters $\xi$ and $m_{X}/m_{\ell}$. This typical degeneracy reveals that consulting tables of constants for the uncertainty of $\a$ (say) determined on the basis of QED-EW theory lacks logical self-consistency when assumptions are revised. 

A related notion that fundamental constants are highly over-determined, and themselves do not depend on new theory variations is also false. When it comes to the highest precision, careful reading will discover that rather few highly- specific data and theory elements with the smallest uncertainties dominate the least-squares fits of published tables. The information is not a secret but still rarely noticed.

\subsubsection{Muonic Lamb Shift}

The muonic Lamb shift theory has been transcribed from Antognini et al\cite{Antognini:1900ns} with additions from Ref. \cite{Jaeckel:2010xx} is \ba &  \Delta E( \alpha, \, \xi, \, m_{X}, \, m_{red}, \, r_{p}) = 
206.0336  \alpha^3/\alpha_{\bullet}^3  - 5.2275 r_{p}^2 \alpha^4/\alpha_{\bullet}^4 \nn \\ & +
 0.0332 + 10^9  (m_{X}^4 \xi)/(2 \alpha m_{red} (1 + m_{X}/(\alpha m_{red}))^4) \ea Here $m_{red}$ is the reduced mass. The $\xi$-dependent term is simply first-order perturbation theory using Schroedinger wave functions and the Yukawa potential, which breaks the degeneracy of the Lamb shift. This unsophisticated calculation is adequate when $\a_{X}$ is sufficiently small. We have also displayed the formula to illustrate how our code is organized to depend on one parameter ($\xi$) when $m_{X}$ is sufficiently large. Random searches missed the favored region many times before it was found and understood. 
 
\subsubsection{Electronic Hydrogen and Deuterium}
\label{sec:transitions}

\begin{small} \begin{table}[ht]

\resizebox{0.83\textwidth}{!}{\begin{minipage}{\textwidth}
	\begin{math}
	\begin{array}{lllll}
	\hline
\text{Transition} & f_{expt} \text{ Hz} & f_{our\:calc} \text{\:Hz} & \sigma_{exp} \text{ Hz}\\
	\hline
	\hline
	\nu_H(2S_{1/2}-8S_{1/2}) & 7.70649350012\times 10^{14} & 7.70649350006\times 10^{14} & 8600 \\
	\nu_H(2S_{1/2}-8D_{3/2}) & 7.7064950445\times 10^{14} & 7.7064950444\times 10^{14} & 8300 \\
	\nu_H(2S_{1/2}-8D_{5/2}) & 7.706495615842\times 10^{14} & 7.706495615680\times 10^{14} & 6400 \\
	\nu_H(2S_{1/2}-12D_{3/2}) & 7.991917104727\times 10^{14} & 7.991917104715\times 10^{14} & 9400 \\
	\nu_H(2S_{1/2}-12D_{5/2}) & 7.991917274037\times 10^{14} & 7.99191727409\times 10^{14} & 7000  \\
	\nu_H(2S_{1/2}-2P_{3/2}) & 9.9112\times 10^9 & 9.9112\times 10^9 & 12000 \\
	\nu_H(2P_{1/2}-2S_{1/2}) & 1.057845\times 10^9 & 1.057846\times 10^9 & 9000 \\
	\nu_H(2P_{1/2}-2S_{1/2}) & 1.057862\times 10^9 & 1.057846\times 10^9 & 20000 \\
	\nu_D(2S_{1/2}-8S_{1/2}) & 7.708590412457\times 10^{14} & 7.708590412336\times 10^{14} & 6900 \\
	\nu_D(2S_{1/2}-8D_{3/2}) & 7.708591957018\times 10^{14} & 7.708591956914\times 10^{14} & 6300 \\
	\nu_D(2S_{1/2}-8D_{5/2}) & 7.708592528495\times 10^{14} & 7.708592528361\times 10^{14} & 5900 \\
	\nu_D(2S_{1/2}-12D_{3/2}) & 7.99409168038\times 10^{14} & 7.99409168032\times 10^{14} & 8600 \\
	\nu_D(2S_{1/2}-12D_{5/2}) & 7.994091849668\times 10^{14} & 7.994091849642\times 10^{14} & 6800  \\
	\nu_D(2S_{1/2}-2P_{3/2}) & 9.91261\times 10^9 & 9.91280\times 10^9 & 300000 \\
	\nu_D(2P_{1/2}-2S_{1/2}) & 1.05928\times 10^9 & 1.05923\times 10^9 & 60000 \\
	\nu_D(2P_{1/2}-2S_{1/2}) & 1.05928\times 10^9 & 1.05923\times 10^9 & 60000 \\
	\hline
	\hline
	\end{array} \end{math}

\end{minipage}}
\caption{The experimental values of electronic hydrogen ($eH$) and electronic deuterium ($eD$) transitions compared to our calculation using the best fit with $\a_{X} \neq 0$. The fit also reproduces the other transitions used in previous QED-EW fits as described in the text within a fraction of the experimental uncertainty. An arbitrary value of $m_{X}=50$ MeV has been used. }

\label{tab:transitions}

\end{table} \end{small}

Electronic hydrogen and deuterium spectroscopy are the most challenging tasks because the theory consists of many dozens of formulas, subsidiary formulas, and numerical parameters. Our attention was initially drawn to the $r_{p-eH}$ determination of Beyer {\it et al} \cite{2013JPhCS.467a2003B} as one of the few independent analyses outside of CODATA compilations. The work develops 14 values of $r_{p}$ and $R_{\infty}$ by solving 14 two-parameter, two-data point fits. Each two-point fit uses the data of the $1S2S$ transition, which is ultra-precise, and the datum of one other transition from a standard $eH$ set considered very reliable. The uncertainty of $r_{p}$ is estimated by repeating the calculation adding and subtracting the experimental errors. As an independent check we reproduced this work entirely.\footnote{We thank Th. Udem for patient explanation of the errors of the 2-point fit procedure used in Ref. \cite{2013JPhCS.467a2003B} and providing computer code to check it.}. We mention this because Figure 1 of Beyer {\it et al} has been widely circulated as ``the standard'' approach to determining $r_{p}$ with $eH$. Each row of the figure shows one of the $r_{p}$ values with its error bars compared to a vertical line for the average value. The procedure is not a global fit, nor designed to compete with one, but instead a demonstration study made with the virtues of simplicity, transparency, and independence. That explains why the uncertainty found for $r_{p}$ is relatively large.

The ``highly reliable'', standard set of precision hydrogen transitions selected by CODATA has been used unchanged for many years in the global QED-EW fits of fundamental constants, with $1S3S$ data added in CODATA 2010\cite{Mohr:2012tt}, henceforth C-10. Our global fit was at first designed simply to reproduce that work, in order to explore the actual uncertainties and correlations. Thus more than a year before we imagined new physics might be relevant we used the existing atomic QED-EW theory to reproduce all those data within a small fraction of experimental uncertainties.  

The computational code\setcounter{footnote}{0}\footnote{The values appearing in Table IX of C10 \cite{Mohr:2012tt} for the $P$-wave parameters $B_{61}$ are erroneous. Since Ref. \cite{Jentschura:2003we} is cited for these parameters, we used the values found in Ref. \cite{Jentschura:2003we}.} is about 30000 characters of Mathematica done with independently written implementations on two different machines. Basic estimates would convert this to about 270,000 characters of C++ code. Validation by line-by-line checking is impossible with independent implementations, so validation was done by fitting data and checking we generate the same numbers, as well as published ones, up to rounding errors. This was done whether {\it including} or {\it excluding} the $1S2S$ transition data from the fits, which was previously thought necessary to obtain sufficient precision. We also reproduced to 13-digit accuracy and better the independent theoretical implementation of level frequencies contributing to transitions, as listed in Table 4 of A. Kramida's review\cite{2010ADNDT..96..586K}, which were obtained from Jentschura {\it et al} \cite{URL}). Note that it is more demanding to compute level frequencies than transitions because many corrections cancel in transitions. The mean difference of the predictions was 65 Hz with a standard deviation of 568 Hz. We also quantify the difference of theory calculations with the ratio $\Delta f_{tt}=(f^{theory-1}-f^{theory-2})/\sigma_{exp}$ computed for each energy level. We use $\sigma_{exp}$ (not estimated theory uncertainties) to avoid theoretical prejudice, and also because the comparison with the experimental uncertainty is what matters in the end. We found $\Delta f_{tt}^{2}< 0.04$ in every case, with a mean for the set of 0.003 and standard deviation of 0.010.  With few exceptions, the C10-selected transitions are simply those with the smallest experimental uncertainties. These transitions are listed with numerous correlation parameters\footnote{We verified that including the input correlations listed in $C10$ for the {\it experimental} data had negligible effects on our QED-EW study: $r_{p}$ was the same within our uncertainty. Indeed the $1S2S$ datum is listed as completely {\it uncorrelated}} and additive corrections discussed in the Appendix. To eliminate a possibility those data are special, we fit the rest of the levels listed in Table 4 of Ref. \cite{2010ADNDT..96..586K} and checked its statements\footnote{Kramida\cite{2010ADNDT..96..586K} discusses 10 cases of calculations differing from experiment by more than $2\sigma$, which all involve $n=3$ or $n=6$ levels with a nearly constant energy shift attributed to systematic experimental error. We also verified those calculations.} In particular, Kramida writes: ``However, one thing can be stated with certainty: the exact agreement of those two ultra-precise $1S2S$ measurements with the QED calculations cannot be considered as a confirmation of the QED theory, because it is the result of the fitting of the fundamental constants based on these (and other) transitions.''  This remark is explained in the Appendix.

 \subsubsection{Our Transitions}

To avoid complicating the proton size puzzle, we also initially restricted attention to $eH$ spectroscopic data, excluding $eD$. Deuterium QED-EW theory involves a change in the reduced mass, a few non-obvious effects of the spin-one deuteron, some changes of computed parameters, and a new charge radius parameter $r_{d}$. Basic nuclear theory predicts $r_{d}^{2} = r_{p}^{2}+ r_{deut}^{2}$, where $r_{deut}$ is a bound state scale which nuclear theory predicts. If this is accepted at sufficient precision, the $eH$ proton charge radius should predict the $eD$ one, and vice versa.  But if the nuclear theory is challenged, the deuteron charge radius becomes another free parameter, which is to be avoided.

The situation changes when the muonic Lamb shift in deuterium becomes experimentally available. Then even if $r_{d}$ is a free parameter, it is over-determined. If the nuclear theory is accepted, it is over-determined twice. In view of the pivotal scientific power of deuteron measurements we report a joint fit to both $eH$ and $eD$ in Table 1 accepting the nuclear theory. This is discussed more in Section \ref{sec:deuts}.

We now explain our selection of 7 transitions each for $eH$ and $eD$ listed in Table \ref{tab:transitions}. In the first place, we fit {\it all} the transitions to within a small fraction of the experimental error bars, except for the $1S2S$. We selected the subset shown to avoid a unduly large number of data skewing the least-squares weight of the rest of the observables. Our 14 transitions are also the entire set not relying on a technique of subtracting fractional combinations of the $1S2S$ transition. The subtraction technique 
is done to cancel out known level-dependent patterns of theory corrections, plus some expected from un-calculated terms. \footnote{We initially used the subtraction technique because others had used it. Except for refining the smallest possible error bars on $\R$, it made no significant difference in the results. One reason to eliminate it is to avoid the need to justify it.} Once again the usefulness of this device depends on the hypothesis. It has been used as a clever way to improve the determination of $\R$ when the QED-EW theory is considered exact. Yet it will hide potential discrepancies if theory is not exact. When considering new physics we wish to discover potential discrepancies, not suppress them.

\subsubsection{Neutron Interactions}

\label{sec:deuts}

Information exists on possible interactions of a new ultra-light boson with neutrons. For coupling constants $g_{e},  \, g_{n}$ the limit of Barbieri and Erickson\cite{Barbieri:1975xy} is $g_{e}g_{n}/4\pi \lesssim 3.4 \times 10^{-11}( m_{X}/MeV)^{4},$ which has been unsurpassed for 40 years\footnote{Note that Ref. \cite{Barger:2010aj} assumes early a coupling to neutrons, which affects limits after that step.}.  

The analysis of Table 1 assumes no new interaction with neutrons. Rather than treat the deuteron charge radius as a free parameter, it is predicted using the global fit value of $r_{p}$ and nuclear theory with $r_{deut}=1.9529$. This is the first test of the model, which could have failed with deuterium. To be fair, nuclear theory is not critical, and the test is mild, because the QED-EW theory fits using $r_{d}$ as a free parameter were known to be consistent with theory. We will also divulge that we explored fitting $r_{d}$ as a free parameter. The best fit value differed from the predicted one by a fraction of a percent with negligible statistical significance.

The agreement of our analysis with $eD$ spectroscopy puts an {\it upper limit} on the size of new neutron interactions. This limit is stronger than Barbieri and Erickson's for $m_{X} \gtrsim 2$ MeV. This fact, plus finding an excellent fit with equal electron, muon and proton couplings, are the basis for us to assume couplings are proportional to electric charge. It is easy to relax that assumption and explore a larger region of allowed parameters.

Using no free parameters, our results predict the deuterium charge radius $r_{d} =2.128$ to be measured independently in the $\mu D$ Lamb shift. This prediction can be done either with $r_{p}$ fit globally, or fit excluding the $eD$ data: See the next Section. The value of $r_{d}$ itself is not new, and e.g. appeared in the 2013 CREMA paper\cite{Antognini:1900ns} projecting future measurements assuming $r_{p} \sim 0.84$, the muonic value. Close to the same preliminary experimental value has been circulating for well more than a year, yet without appearing in print. Agreement is non-trivial. A theory of a new muon-specific interaction would have a proton charge radius close to the QED-EW value $r_{p-QED} \sim 0.878$. The value of $r_{d}$ would agree in $eD$ but not in $\mu D$. Moreover, the muonic deuterium results have still not been officially released as we write this paper. It is ironic that all the information to make a prediction existed as early as 2010, upon discovery of the proton-size puzzle in muonic hydrogen. But at that time we did not understand the importance of the puzzle for the interconnections between the fundamental constants.

\subsection{Parameter Ranges, $\chi^{2}$ Budget, and Analysis Variations} 

We checked and extended our results extensively by re-fitting data with and without different classes of observables.

Table \ref{tab:parameterTable} lists parameters and their uncertainties obtained from the full fit and fits removing particular data classes for the arbitrary value $m_{X}=50$ MeV. In some cases the effects of removing an observable are easy to anticipate. For example, removing $\mu H$ causes $r_{p} \ra 0.88$ exactly as found in previous work removing them. The small uncertainty of the $\mu H$ datum causes a rather small uncertainty in $r_{p}$ determined using it. In other cases parameters vary significantly due to non-obvious interplay between fundamental constants.

\begin{table}[ht]
  \centering
\resizebox{.83\textwidth}{!}{\begin{minipage}{\textwidth}
\begin{math}
\begin{array}{ccccccc}
 \hline
 \text{Omit} & \chi ^2(\lambda_{c}) & \chi ^2 (\mu H) & \chi ^2 (a_e) &
   \chi ^2 (a_{\mu}) & \chi ^2 (eH) & \chi ^2 (eD) \\
   \hline
   \hline
 \text{none} & 1.6 & 0.00084 & 1.5 & 0.18 & 6.8 & 4.2 \\
 \text{$\lambda_c$} &  \text{--} & 0.00068 & 4.\times 10^{-9} & 0.0030 & 6.8 & 4.2 \\
 \text{$\mu$H} & 1.6 &  \text{--} & 1.5 & 0.23 & 3.3 & 3.5 \\
 \text{$a_{e}$} & 4.4\times 10^{-9} & 0.00078 &  \text{--} & 0.0030 & 6.8 & 4.2 \\
 \text{$a_{\mu}$} & 0.024 & 0.00087 & 0.023 &  \text{--} & 7.4 & 4.3 \\
 \text{$a_{e}$, $a_{\mu}$} & 6.7\times 10^{-8} & 2.0\times 10^{-12} &  \text{--} &  \text{--} & 3.3 & 3.5 \\
 \text{$\mu$H, $a_{\mu}$} & 6.7\times 10^{-8} &  \text{--} & 9.5\times 10^{-13} &  \text{--} & 3.3 & 3.5 \\
 \text{eH} & 1.5 & 5.6\times 10^{-6} & 1.4 & 0.22 &  \text{--} & 4.2 \\
 \text{eD} & 1.6 & 0.00057 & 1.5 & 0.18 & 6.8 &  \text{--} \\
 \text{eH, eD} & 0.0 & 3.1\times 10^{-17} & 9.7\times 10^{-14} & 2.5\times 10^{-15} & \text{--} & \text{--} \\
\hline
\hline
\end{array}
\end{math}

\end{minipage}}
\caption{Contributions to $\chi^{2}$ at a reference point $m_{X}=50$ MeV. $\Delta \chi^{2}$ is the difference in $\chi^{2}$ between the null model ($\a_{X}=0$) and the best fit. Also shown are the contributions with different observables omitted. Fits are made with the arbitrary value $m_{X}=50$ MeV. The columns of $\chi^{2}$ and $\Delta \chi^{2}$ are the same as those in Table \ref{tab:parameterTable}, hence not repeated. }

\label{tab:chibudget}

\end{table}

The budget of $\chi^{2}$ for each class is shown in Table \ref{tab:chibudget}. The value of $\chi^{2}$ and $\Delta \chi^{2}$ are the same for each row as Table\ref{tab:parameterTable}, hence not repeated. Over the range of $m_{X}$ in the favored region each type of contribution is close to statistical expectations for the Birge ratio, commonly expressed with $\chi^{2}/dof$. The contribution to $\chi^{2}$ of $a_{e}$ has a local maximum of three at $m_{X} \sim 20$ MeV. This is still acceptable in view that every analysis of physics beyond the Standard Model allows a minimum $2 \sigma$ variation in the experimental value of $a_{e}$. If $a_{e}$ is dropped from the analysis our best-fit parameters for large $m_{X}$ are hardly affected, except for degrading the precision of $\a$. The rapid variation of $\chi^{2}$ as a function of $m_{X}$ is largely due to sensitive dependence of $a_{\mu}$ (immersed in the global fit) to $m_{X}$ in the range $m_{X} \lesssim m_{\mu}/2$.

Removing the deuterium data causes negligible changes in parameters: All remain within the uncertainties given in Table 1. Table \ref{tab:DeutoCompo} shows the parameters assuming $eH$ data only. Because of this, our analysis using $eH$ can {\it predict} the body of $eD$ data to within fractions of the experimental uncertainties. That is impressive but dominated by the fact that conventional QED-EW theory has high predictive power once $\R$ and $r_{p}$ are determined. 

Removing the muonic Lamb shift data significantly changes fit parameters. The value of $r_{p}$ goes to 0.88, as found in previous QED-EW work (C10) excluding $\mu H$ and $a_{\mu}$. A region of $m_{X} \sim 45$ MeV is favored, which is related to the $(\a_{X}, \, m_{X})$ range previous long determined capable of fitting $a_{\mu}$.

The changes $\Delta \chi^2$ shown in Table \ref{tab:chibudget} indicates the $\a_{X}  \neq 0$ model case is highly favored in all cases, except when 
$a_{\mu}$ is removed. Assessing this needs to balance the penalty of the new model using an extra parameter $\xi$, versus the penalty for excluding data the QED-EW theory does not fit. The status of $a_{\mu}$ should become more clear with the upcoming Fermilab experiment. We found it interesting to accept $r_{p}=0.84$ and re-evaluate fundamental constants in the QED-EW null model. That exercise predicts $a_{\mu} =  0.0011659183957 $, a $15.7 \sigma$ discrepancy with the Standard Model. The oft-quoted 3.9$\sigma$ discrepancy come from using a fit to fundamental constants excluding $a_{\mu}$ and $\mu H$ data entirely. 

Since the value of $a_{\mu}$ may change with the Fermilab experiment, it would be interesting to explore the range of $a_{\mu}$ over which either the Standard Model or a one-parameter new model would be compatible. In this regard we note that studies of new physics confronting the electron anomalous moment $a_{e}$ invariably use the experimental value minus $2\sigma_{ae-exp}$. That is because new interactions make a positive contribution at one-loop order, while the QED-EW theory prediction is already larger than the experimental one. There are no experimental consistency checks on $a_{e}$ outside one group's measurement\cite{2008PhRvL.100l0801H}, so the practice of adjusting the data seems acceptable. Nevertheless our fits are done with $a_{e}$ set at the value reported.

\section{Exclusion Limits} 
\label{sec:limits}

The first question on exclusion concerns the {\it spin} of the exchanged boson $X$. We left the spin undetermined in making fits to a generic Yukawa interaction at low momentum transfer. A spin-0 interaction between fermions is characterized by $\gamma_{5}$  (pseudoscalar) or $1$ (scalar) vertices. The $\gamma_{5}$ form produces a derivative interaction via chiral Ward identities. In a field theory a fundamental $\gamma_{5}$ interaction also needs to contend with knotty ultraviolet consequences of chiral anomalies. This leaves a scalar interaction. Under broad conditions a scalar interaction between identical particles or antiparticles is attractive\cite{Peskin:1995ev}. Our interaction is {\it repulsive}, ruling out spin-0 for interactions scaling like electric charge.\footnote{Despite lore to the contrary, we have not seen a correct proof that scalar interactions are attract with the most arbitrary coupling assignments. Totally arbitrary couplings would greatly increase our parameter space, contrary to the goal we have set.} This leaves a spin-1 exchange as the main candidate.

Limits on a new light vector boson coupling to electrons differ significantly if $m_{X} \leq 2m_{e}$, preventing decay to $e^{+}e^{-}$ pairs, compared to otherwise. Our fits have identified the regime $m_{X} >2 m_{e}$ to be the region of interest. A community concentrating on {\it dark photon} models\cite{Fayet:2006sp, Pospelov:2008zw, Essig:2013lka} has led to compilation of experimental bounds on light vector boson in a model with a coupling constant $g =\epsilon e$, for electric charge $e$. In most renditions the parameter $\epsilon$ measures kinetic energy mixing of the usual $U(1)$ and a new $U(1)'$ gauge boson. That is by no means the unique road to a new interaction. We did not begin with the model, which is by no means the unique road to a new interaction, and in fact the sign of our coupling is the opposite of that predicted by simple kinetic mixing.  Nevertheless, the parameter limits developed with dark photon models have important information. Bounds are commonly expressed in terms of $\epsilon^{2}$, because most experiments are not sensitive to the sign of the coupling. We can then transcribe $\epsilon^{2} \ra \a_{X}$, subject to the understanding that our analysis is done ``bottom up'' empirically with a parameter $\a_{X}$ fit to data, for which we have no other information.

Figure \ref{fig:limitplot} shows our region of best-ft in the $(\a_{X} , \, m_{X})$ plane superposed on a plot adapted from Ref. \cite{Essig:2013lka}. The favored region comes from finding the curve $ \chi^{2}(\a_{X} , \, m_{X})=minimum$ with $\Delta \chi^{2}>4$ ($m_{X}=20$ MeV) ranging to $\Delta \chi^{2}>13$ ($m_{X}=200$ MeV). The region can be extended indefinitely for larger $m_{X}$ through kinematic dependence on $\xi =\a_{X}/m_{X}^{2}$. We do not determined an upper limit on $m_{X}$. For $m_{X} \gtrsim 200 $ MeV the minimum $\chi^{2}$ varies so slowly no significant resolution of $m_{X}$ occurs. Once the $ \chi^{2}(\a_{X} , \, m_{X})=minimum$ is found, the favored region is defined by varying it by $2\sigma_{\alpha_{X}}$, where $\sigma_{\alpha_{X}}$ is the uncertainty of $\a_{X}$ point by point.  

Other colored regions in Fig. \ref{fig:limitplot} show where previous work has excluded dark photons, {\it subject to certain assumptions needed in those analyses}. We briefly discuss the cases where our favored region crosses a potentially excluded region:

$\bullet$ The BaBar exclusion region is based on missing momentum in $\Upsilon$ decays to invisible final states. Assuming a universal coupling to all quark generations, which our study cannot in principle determine, one can transcribe $\epsilon^{2} \ra \a_{X}$. With that assumption the region where our ``favored'' region crosses the BaBar region appears to be ruled out. Any model coupling to $b$-quarks smaller than the light quarks will weaken or nullify the limit. Since we have not constructed a model with group representations predicting $b$-couplings, we let the favored region cross the BaBar region. Nothing from our study but perturbative consistency determines an upper limit on $m_{X}$. The graphics have not been extended to high masses because $\xi$ dependence makes extrapolation straightforward.

$\bullet$ The A1 exclusion region\cite{Merkel:2014avp} confirmed and superseded the WASA\cite{Adlarson:2013eza} and HADES\cite{AgakishievHADES2013} limits also shown. The experiment hinges on decay to $e^{+}e^{-}$ pairs whose invariant mass spectrum is measured. The bounds assume the branching ratio of $X$ to $e^{+}e^{-}$ is unity. By making that assumption the electron interaction of our model is constrained and potentially ruled out where it crosses the A1 region. It is also well known that such bounds are weakened or nullified in models decaying preferentially to invisible particles, such as neutrinos or dark matter candidates. Nothing in our data analysis excludes that possibility.

$\bullet$ The region of $m_{X} \lesssim 10$ MeV is severely constrained by E774 (\cite{Bjorken:2009mm}) shown at the left edge of the plot, and many other studies relevant to smaller $m_{X}$ listed in Ref. \cite{Essig:2013lka}. The full analysis value of $\Delta \chi^{2}$ value we find happens to not be significant in a region $m_{X} \lesssim 10-20$ MeV. The physics and bounds of the region $m_{X}< 1$ MeV are quite different and generally difficult to reconcile between $a_{e}$, $a_{\mu}$ and $\mu H$. Our search setting the coupling of electrons to zero recovered the parameter region near $m_{X}\sim 1$ MeV previously found in Refs. \cite{Barger:2010aj, TuckerSmith:2010ra}.

$\bullet$ The region of 20 MeV$ \lesssim m_{X} \lesssim 40$ MeV and $\a_{X}\lesssim 7 \times 10^{-5} $ is open and at the same time favored. A substantial portion of this region will be explored by upcoming or proposed new experiments. The list includes BDX, DarkLight, HPS, VEPP-3, APEX-2 at Jlab, new experiments at MESA (Mainz), BelleII (Kek), MU3E (PSI), Seaquest at FNAL, and the LHC\cite{Ilten:2016tkc}. Any of these experiments might potentially discover $X$ in or near our favored region.

We caution that our review of the mass range should not be interpreted as a final determination of $m_{X}$. One can certainly make a well constrained prediction subject to assumptions. There remains to explore the increased range of parameters from actually varying the experimental inputs by a few units of their reported uncertainties. We have not yet investigated this beyond finding $\Delta \chi^{2}$ increases about 2 units across the favored region when all experimental uncertainties are doubled. 

\section{Discussion}

We have compared fits to high precision experimental data using the Standard Model and a generic model adding a low mass, weakly interacting boson $X$. The data includes the electron anomalous moment, electronic hydrogen and deuterium spectroscopy, the electron Compton wavelength, plus the muonic Lamb shift and muon magnetic moment which have been excluded from previous high-precision global fits. Logical consistency demands globally fitting the fundamental constants to the new theory when the new theory is used. A conventional $\chi^{2}$ statistic rules out the Standard Model compared to the new one by about 13 units of $\Delta \chi^{2}$ at the reference point of boson mass $m_{X} =50$ MeV. The new favored region of fundamental constants happens to agree within uncertainty with previous determinations of $\a$ and $\lambda_{c}$ while disagreeing with $\R$ by $2-3\sigma$. That is quite acceptable, because previous determinations of $\R$ and its uncertainty referred to a different theory. Other experimental observables are less restrictive and consistent. No upper limit on $m_{X}$ is determined. The minimal-universal solution is not restricted to any particular Lagrangian density, but appears to favor a spin-1 intermediate boson. Fits have been conducted using one universal coupling between $e$, $\mu$ and $p$, finding values of $\xi=\a_{X}/m_{X}^{2} \sim 1.2 \times 10^{-11}$, corresponding to $\a_{X} \sim 3\times 10^{-8}$ at $m_{X}=$ 50 MeV.  The range of $m_{X} \gtrsim 50$ MeV can be excluded if an assumption is made that $X$ decays with 100\% branching ration to $e^{+}e^{-}$, otherwise not. The range of $20 \text{ MeV} \lesssim m_{X} \lesssim 50 \text{ MeV}$ is not excluded by current limits, while inside the favored parameter region of the new model. A number of approved or planned upcoming experiments can confront the new model in the favored parameter region. 

The minimal-universal solution is unconventional, and unexpected, on the previous assumptions that new interactions should have been more visible in electron-based observables than muon-based ones. That is true, but the agreement of certain electron-based observables is nearly circular due to constants the observables dominate in fits. A global fit to all the constants is necessary to explore the effects. The minimal-universal solution finds the true proton charge radius $r_
{p} \sim 0.84$ is very close to the one determined by muonic hydrogen experiments. There are no free parameters in a prediction of the muonic deuterium charge radius, whose experimental measurement is expected to be announced soon. 

The universal nature of the interaction makes possible many tests that a muon-specific interaction could not confront. Spectroscopic tests include measuring more transitions in muonic hydrogen, detuerium and helium. Electronic hydrogen Rydberg states with $n>>1$ will appear
to indicate two different Rydberg constants. The model predicts effects that should be observable in positrionium, muonium ($e^{-}\mu^{+}$ and $e^{+}\mu^{-}$) and true muonium ($\mu^{+}\mu^{-}$). Depending on $m_{X}$, the trend is that QED-EE theory will disagree with positronium while agreeing with true muonium, due to the relatively more significant effects of a light interaction on electrons. At the momentum transfer of existing experiments $\mu^{\pm} p$ and $e^{\pm}p$ scattering should both find the same apparent charge radius. The pole singularity of $X$ is too small and too close to zero momentum transfer to be resolved with current methods, but might be observable in experiments dedicated to ultra-small momentum transfers. We are optimistic about the prospects for discovery. 

\medskip
{\bf{Acknowledgments:}} We thank Randolf Pohl, Thomas Udem, Graham Wilson, Ron Gilman, Abni Soffer, Roger Barlow, Louis Lyons, Doug Higinbotham, Cynthia Keppel, Stan Brodsky, KC Kong, Doug McKay, Greg Adkins, and Michael Eides for helpful information, discussions or suggestions.

\section{Appendix: Avoiding Unnecessary Sensitivity} 

A basic principle of data analysis hold that no result should be unduly sensitive to procedural decisions, or if there is high sensitivity, it should be understood and divulged. The importance of the issues demand that procedures also be direct, transparent and
reproducible by others. This is why our analysis considers the most simple possible least-squares fit using experimental uncertainties. We now discuss the theoretical uncertainties postponed to this Section.

There are no universal rules for incorporating estimated theory uncertainties in data analysis. Barlow\cite{Barlow:2002yb} has explained theory uncertainty is an intrinsically Bayesian issue. We explored several approaches. The method called ``chi-squared with pull'' adds new parameters $\delta_{j}$ to the theory, replacing$ (d_{j}-t_{j}(\theta_{\ell}) )^{2} \ra  (d_{j}-t_{j}(\theta_{\ell}) +\delta_{j})^{2}$ in Eq. \ref{chidef}. Additional terms are also added to $\chi^{2}$ to regulate how much $\delta_{j}$ can vary. The hypothesis that $\delta_{j}$ are normally distributed about zero with estimated uncertainties $\sigma_{\delta-j}$ adds $\sum_{i} \, \delta_{j}^{2}/\sigma_{\delta-j}^{2}$ to $\chi^{2}$. The results then depend on $\sigma_{\delta-j}$, which are essentially free parameters representing one's belief in the theory. The Appendix of Ref. \cite{Stump:2001gu} reviews this and warns that fitted outputs can be unexpectedly sensitive to the $\sigma_{\delta-j}$.

The method tends to punish high confidence in theory, and reward low confidence, somewhat counter-intuitively. If the theory is not trusted, then $\sigma_{\delta-j}$ are large, allowing the additive parameters to shift the theory and fit the data better. However the range of theory parameters fitting within a given confidence level is also increased, downgrading parameter resolution. High confidence in theory is represented by small $\sigma_{\delta-j}$ that prevents additive parameters from helping the theory. Like all Bayesian procedures the results depend on one's beliefs about the theoretical uncertainties $\sigma_{\delta-j}$, known as priors. The process of fitting the $\delta_{j}$ can be bypassed (in Bayesian terms, concealed) if one marginalizes over the distribution of priors. For a normal distribution that replaces $\sigma_{exp-i}^{2} \ra \sigma_{exp-i}^{2}+ \sigma_{\delta-i}^{2}$ in the denominators of $\chi^{2}$. ``Add theory and experimental errors in quadrature.'' The formula automates a rule that if theory uncertainties are {\it sufficiently small} compared to experimental ones, they have no effect. 

Almost by definition, theoretical uncertainties must be smaller than experimental ones to discover experimental anomalies. (When the opposite happens, the theory is inadequate to confront the data, and discrepancies do not become anomalies.) The decision that anomalies exist, at least for discussion, takes as a starting point that theory errors are not the leading candidate for explanation. As consistent, almost all of the data and theory elements of our study have been repeatedly examined to rule out an important role for theoretical uncertainty. 

For example, the theory of the muonic Lamb shift\cite{Pachucki:1996zza} is beautifully simple, compared to electronic Lamb shift. The proton size contribution is {\it ten million times larger than in electronic hydrogen}, and almost all of of it comes from {\it first order} perturbation theory. The muonic Lamb shift is theoretically robust, and calculations are complete. Higher order corrections make small contributions, and they have been calculated from first principles. Theory uncertainties have already been combined with experimental ones in the reported uncertainties we use. 

The theory of the electron anomalous moment is quite difficult. It has only been computed to the highest precision by one group, and significant mistakes have been found in the past. Yet we have no insight to irevise the estimated theoretical uncertainty. Any decision by us to increase it might be perceived as an unfair bias making the discrepancies easier to explain. That contradicts our study, so we have no option but to accept the experimental uncertainty used by the community, which is larger.

The theory of electronic hydrogen and deuterium is extremely complicated. The estimated theoretical uncertainties of $\a log\a$ series expansions do not always agree with calculations done after the estimates. Higher order terms are not reliably of order $\a/\pi$ relative to lower order ones. Here again we have a dilemma that if we increased theory uncertainties it would unfairly bias our study. Fortunately there are consistency checks. Almost all of the electronic hydrogen and deuterium spectra are all fit to within a fraction of the experimental uncertainty with $\chi^{2}/dof <1$, exactly as consistent with the estimated theory uncertainties. To explore this in more detail, we did repeat the electronic hydrogen fits including additive corrections and correlations mentioned earlier, and used in C10, to verify they have negligible effects. This exercise was redundant, because the outcome can be found analytically and always happens when estimated theory uncertainties are sufficiently small, {\it and} the theory fits the data without additive parameters. We did the work because we anticipated a demand to demonstrate it. We decided on the simpler and more transparent procedure presented in the text for the virtue of demanding a minimum to explain it, justify it, and for allowing no perception of bias favoring the theory.

In summary, our analysis appearing to ignore theory errors is the most conservative treatment of
theory uncertainty for the purpose of our analysis. Any method increasing theory uncertainty would make
explaining the experimental anomalies easier. It would improve fits by decreasing $\chi^{2}$ while decreasing parameter resolution.

\subsection{The Exceptional Datum}

We turn to the sole exception to all of the above, which is the $1S2S$ transition of $eH$. The experimental uncertainty of this transition\footnote{The $1S2S$ uncertainty was 35 Hz in C10} is only 10 Hz, compared with the transition's overall value of $2.46 \times 10^{15}$ Hz, putting it in a class of the most relatively precise measurements of all time. The rest of the standard hydrogen dataset (14 transitions listed in C10) have $\sigma_{j}^{2}$ ranging from $4.1 \times 10^{7} \, Hz^{2}$ to $5.8 \times 10^{8} \, Hz^{2}$. The mean value of $\sigma_{j}^{2}=148,400 \, \sigma_{1S2S}^{2}$. The relative weight of one datum 148,000 times more important than others signals an extreme sensitivity of $\chi^{2}$ to the $1S2S$ transition. Due to one ultra-precise point, the minimum $\chi^{2}$ for a simple least squares fit (no pull term corrections) including the $1S2S$ datum would be 122,500 units, based on a minimal 3.5 kHz theory error for that point. It would seem a great accomplishment to fit the $1S2S$ level.

The theory however has two parameters $r_{p}$ and $\R$ that can be freely varied. It is always possible to satisfy one constraint -- namely fitting the $1S2S$ to arbitrary precision --with two parameters. Before the muonic Lamb shift disturbed the scene, those parameters had significant freedom, because other experiments determined them much less precisely. The analysis constraint of fitting the $1S2S$ data with QED-EW theory and linearizing in $r_{p}$ is \ba r_{p, \,1S2S} \sim 0.877 + & 1.05 \times 10^{9}\delta R_{\infty}/ \R^{\bullet},  \label{one}. \ea The relation is good for $10^{9}\delta R_{\infty}/ \R^{\bullet}<<1.$ Eq. \ref{one} will be called the ``the artificial $1S2S$ degeneracy line'', or ``$1S2S$ correlation''. It refers to an artificial proton size parameter $r_{p, \,1S2S}$ deduced from one data point and no other data. From this artificial relation, and nothing more, one can find the experimental uncertainty of $r_{p, \, 1S2S}$ given the uncertainty of $\R$, and vice-versa. When other data of current precision are added, their weight in $\chi^{2}$ is far too small to change the $1S2S$ correlation, which controls the subsequent analysis.

Continuing, Eq. \ref{one} comes from setting $data \equiv theory$ for the $1S2S$, so it is subject to the uncertainty of the theory. {\it If the theory uncertainty were small compared to 10 Hz}, the $1S2S$ degeneracy line would be a reliable statement. {\it Yet the most optimistic estimates of $1S2S$ theory uncertainty are huge compared to 10 Hz}. A few years ago the $1S2S$ theory uncertainty was listed as about 20 kHz. Ref.\cite{2005PhRvL95p3003J} lists a number translating to 23 kHz uncertainty if the $1S2S$ correlation were not used in the analysis. A few papers revised the estimated uncertainty down to several kHz when a parameter called $B60(1S)$ was partially calculated by two groups. The different groups reported $B60$ as contributing -6.2 kHz or -12.7 kHz, a 100\% difference which might be a starting point\footnote{Note the calculations themselves are incomplete, and disagree by many units of their estimated uncertainties. Readers can consult the literature to find how other uncertainties have been estimated. } for the estimating theory uncertainty. Specifically, the values are 620 or 1270 times the experimental uncertainty\cite{2003PhRvL..91k3005P, Yerokhin:2003pq, 2005PhRvA..71d0101Y, 2005NIMPB.235...36Y, 2007CaJPh..85..521Y}. 

Given that the theory uncertainty is much larger than the experimental one, any analysis using the $1S2S$ datum becomes highly sensitive to how the theory uncertainty is handled. One obviously has the freedom to interpolate between the artificial $1S2S$ degeneracy line, ignoring the fact its information is unreliable, to downgrading the weight of the $1S2S$ to no weight. We explored this with the additive correction method. Adjusting the regulator of the additive correction within independent uncertainty estimates (of order 20kHz) was enough to double the error bars of the QED fit to $\R$ and $r_{p}$.

We decide to dispense with additive corrections, fit data without the $1S2S$ constraint, and predict the $1S2S$ transition from the rest of the $eH$ data. This agreed with experiment within 3.5 kHz. We found the same results in the global fit including $\a_{X}$. The agreement of our fit within the smallest of all estimated theory uncertainties is acceptable, if perhaps fortuitous. With the agreement, we retrospectively constructed a pull term regulator $\sigma_{\delta -1S2S} =3.5$ kHz knowing it would yield the same result, which we specifically checked by redoing analysis including the $1S2S$. We have not reported fits on that basis because the appearance of $\sigma_{exp}^{2}+\sigma_{theory}^{2}= 10^{2} \, Hz^{2} + 3500^{2} Hz^{2}$ in the denominator of $\chi^{2}$ would certainly raise questions about the arbitrary number 3500. If our text suggests we'd want to defend it, we won't. We fit the data well enough without any maneuvering, and making data fits more elaborate than they need to be is not generally productive. 

The $1S2S$ datum has a strong influence on the current experimental puzzles whether or not physics beyond the Standard Model is considered. When the $1S2S$ transition is omitted, the QED-EW determination of $\R$ and $r_{p}$ are  $r_{p } = 0.87 \pm 0.01 \, fm$ and $\R =1.097373156851\times 10^7 \pm 8 \times 10^{-5} \, m^{-1}$. When this information is used to asses the proton size puzzle, the $7\sigma$ discrepancy of the muonic Lamb shift becomes a $3-3.5\sigma$ discrepancy. That is quite a change in  confidence level, because $3\sigma$ effects ($10^{-3}$ P-value) occur much more often than $7\sigma$ effects ($10^{-12}$ P-value). We need to divulge this because some might find the information sufficient reason to re-assess the proton size puzzle.

To conclude, the attempt to use any ultra-precise data whose theoretical uncertainty greatly exceeds its experimental uncertainty leads to a Bayesian dilemma. No resolutions exist where results {\it do not} depend exquisitely on prior beliefs and arbitrary analysis decisions. Since there are no absolutely right or wrong data analysis procedures, it is certainly possible to use the ultra-precise $1S2S$ datum in many ways. However if using it produces a significant change in results, the change will be highly sensitive to subjective decisions about theory uncertainties, which tend to be contentious. If using the point does not produce significant changes, the datum can be omitted from the analysis, simplifying everything. This explains our decision to omit the $1S2S$ transition from the analysis reported.

\begin{landscape}
\begin{table}[htbp]
	
	\resizebox{0.73\textwidth}{!}{\begin{minipage}{\textwidth}
\begin{math}
\begin{array}{|c|c|c|c|c|c|c|c|c|}
\hline
\multicolumn{1}{|c}{ }&\multicolumn{2}{|c}{ \delta  \R/\R^{\bullet}}&\multicolumn{2}{|c}{  r_p \, [\text{$fm$}] }&\multicolumn{2}{|c}{ \delta \alpha/\alpha^{\bullet} }&\multicolumn{2}{|c|}{ \alpha_X/\alpha} \\
\hline
 m_X \, [\text{$ MeV/c^2$}] & full \, \, fit & without \, \, D & full\, \, fit & without \, \, D & full\, \, fit & without \, \, D & full\, \, fit & without \, \, D  \\
 \hline
 \hline 
10 & 4.5(7.2)\times 10^{-10} & 4.4(7.2)\times 10^{-10} & 0.84099(36) & 0.84098(36) & -2.3(3.6)\times 10^{-10} & -2.3(3.6)\times 10^{-10} & 1.0(2.4)\times 10^{-7} & 1.0(2.4)\times
   10^{-7} \\
 25 & 8.7(5.2)\times 10^{-10} & 8.7(5.2)\times 10^{-10} & 0.84130(31) & 0.84130(31) & -4.4(2.6)\times 10^{-10} & -4.4(2.6)\times 10^{-10} & 2.10(80)\times 10^{-6} & 2.09(80)\times 10^{-6}
   \\
 50 & 6.1(4.3)\times 10^{-10} & 6.1(4.3)\times 10^{-10} & 0.841130(27) & 0.841130(27) & -3.1(2.1)\times 10^{-10} & -3.1(2.1)\times 10^{-10} & 4.8(1.3)\times 10^{-6} & 4.8(1.3)\times
   10^{-6} \\
 100 & 4.3(4.1)\times 10^{-10} & 4.3(4.1)\times 10^{-10} & 0.84101(26) & 0.84101(26) & -2.2(2.1)\times 10^{-10} & -2.2(2.1)\times 10^{-10} & 9.4(2.4)\times 10^{-6} & 9.4(2.4)\times
   10^{-6} \\
 150 & 3.8(4.1)\times 10^{-10} & 3.8(4.1)\times 10^{-10} & 0.84097(26) & 0.84097(26) & -2.0(2.0)\times 10^{-10} & -2.0(2.0)\times 10^{-10} & 1.49(38)\times 10^{-5} & 1.49(38)\times
   10^{-5} \\
 200 & 3.6(4.1)\times 10^{-10} & 3.6(4.1)\times 10^{-10} & 0.84095(26) & 0.84095(26) & -1.9(2.0)\times 10^{-10} & -1.9(2.0)\times 10^{-10} & 2.17(56)\times 10^{-5} & 2.17(56)\times
   10^{-5} \\
 300 & 3.4(4.1)\times 10^{-10} & 3.4(4.1)\times 10^{-10} & 0.84094(26) & 0.84094(26) & -1.8(2.0)\times 10^{-10} & -1.8(2.0)\times 10^{-10} & 3.9(1.0)\times 10^{-5} & 3.9(1.0)\times
   10^{-5} \\
\hline
\hline
\end{array}
\end{math}

\end{minipage}}
\caption{Comparing the full fit to the full fit excluding electronic deuterium. }

\label{tab:DeutoCompo}

\end{table}
\end{landscape}

\section{References} 

  \bibliographystyle{model1a-num-names} 
  \bibliography{bibtest4JM}

\end{document}